\documentclass[aps,prb,superscriptaddress,twocolumn]{revtex4}

\usepackage{bm}
\usepackage{amsfonts}
\usepackage{amsmath}
\usepackage[dvips]{graphicx}

\begin{document}



\title{Entanglement in quantum critical spin systems}

\author{Tommaso Roscilde}
\affiliation{Department of Physics and Astronomy, University of
Southern California, Los Angeles, CA 90089-0484}
\author{Paola Verrucchi}
\affiliation{\mbox{Istituto Nazionale per la Fisica della Materia, UdR Firenze,
Via G. Sansone 1, I-50019 Sesto F.no (FI), Italy}}
\author{Andrea Fubini}
\affiliation{\mbox{Istituto Nazionale per la Fisica della Materia, UdR Firenze,
Via G. Sansone 1, I-50019 Sesto F.no (FI), Italy}}
\affiliation{\mbox{Dipartimento di Fisica dell'Universit\`a di Firenze,
Via G. Sansone 1, I-50019 Sesto F.no (FI), Italy}}
\author{Stephan Haas}
\affiliation{Department of Physics and Astronomy, University of
Southern California, Los Angeles, CA 90089-0484}
\author{Valerio Tognetti}
\affiliation{\mbox{Istituto Nazionale per la Fisica della Materia, UdR Firenze,
Via G. Sansone 1, I-50019 Sesto F.no (FI), Italy}}
\affiliation{\mbox{Dipartimento di Fisica dell'Universit\`a di Firenze,
Via G. Sansone 1, I-50019 Sesto F.no (FI), Italy}}
\affiliation{\mbox{Istituto Nazionale di Fisica Nucleare, Sez.
di Firenze, Via G. Sansone 1, I-50019 Sesto F.no
(FI), Italy}}

\date{\today}
\maketitle

\noindent
Contributed paper to the conference 
\emph{Macroscopic Quantum Coherence and Computing}, Naples, June 2004.

\section{Introduction}
 One of the most striking aspects of quantum coherence
in a quantum many-body system is the occurrence of 
{\it entanglement}, namely the realization of a 
superposition of many-body states that cannot
be factorized into a product of single-particle wave 
functions. An entangled state possesses correlations
that cannot be accounted for by classical-like 
quantities; for instance an entangled state might
not show any form of classical order, and nonetheless be
at the same time strongly correlated. 
The possibility of a local description
of such state is partially or completely lost, 
depending on the degree of entanglement contained
in the state. In particular, the non-local nature 
of special collective 
quantum states is the fundamental ingredient that
allows quantum communication protocols and quantum 
computation algorithms \cite{NielsenC00} to outperform their 
classical counterparts.

 The idea of entanglement as a resource naturally
demands a systematic investigation of which quantum
many-body systems are able to display sizable 
entanglement in a controllable way. An intriguing 
perspective is that pure-quantum correlations
are strongly enhanced when a system undergoes a 
quantum phase transition (QPT) \cite{Sachdev99} 
analogously to what 
classical correlations do at a thermal phase 
transition. Indeed quantum fluctuations show up 
at all length scales at a quantum critical point.
In what sense quantum correlations, and thus 
entanglement, 'diverge' at a QPT has been the 
subject of investigation in several recent 
studies \cite{OsborneN02,Osterlohetal02,
Vidaletal03, Verstraeteetal04}, although the 
resulting picture is still controversial. In this work
we consider a quite general $S=1/2$ quantum
spin system displaying a field-induced QPT 
in arbitrary dimensions. In particular,
we discuss how the behavior of entanglement at and 
around the quantum critical point shows
strong features providing new insight in the 
drastic change of the system's ground state under the 
effect of strong quantum fluctuations.

\section{The model}
 
 A very general example of a quantum phase transition 
 in spin models is offered by the 
 the antiferromagnetic {\it XYZ model} in a field.
 The model Hamiltonian reads:
\begin{equation}
{\hat{\cal H}} =
J \sum_{\langle ij \rangle} \Big[ \hat{S}^x_i\hat{S}^x_{j}
+ \Delta_y \hat{S}^y_i\hat{S}^y_{j} 
+ \Delta_z \hat{S}^z_i\hat{S}^z_{j} - \frac{2h}{\rm z} \hat{S}^z_i \Big]
\label{e.XYZhz}
\end{equation}
where $J>0$ is the exchange coupling, the sum $\langle ij \rangle$ 
runs over the nearest neighboring sites of a bipartite lattice with 
coordination number ${\rm z}$, and 
$h\equiv g\mu_{\rm B} H/J$ is the reduced magnetic field.
In the following we perform the canonical transformation 
$\hat{S}^{x,y}_i \to (-1)^{i} \hat{S}^{x,y}_i$ on Eq. (\ref{e.XYZhz}),
so that the relevant correlations along the $x$ and $y$ axes
are ferromagnetic.   
The parameters $0 \leq \Delta_y, \Delta_z \leq 1 $ control 
the anisotropy 
of the system. In the most general case of $\Delta_y\neq 1$
the Zeeman term in Eq.(\ref{e.XYZhz}) does not commute
with the rest of the hamiltonian. This property is 
at the core of the field-driven quantum phase transition occurring
at a critical field $h_{\rm c}(\Delta)$, which separates a
Neel-ordered phase ($h\leq h_{\rm c}$)
 from a partially-polarized disordered phase ($h>h_{\rm c}$).
 When $h\leq h_{\rm c}$ the field favors long-range magnetic
 order along the $x$-axis. This order disappears at
 the critical field $h_{\rm c}$, where magnetic correlations
along $x$ become short-ranged, while quantum
fluctuations prevent the spins from being fully polarized along
the field \cite{KurmannTM82,Dmitrievetal02,Cauxetal03}.

 The case $\Delta_z = 0$ reproduces the
XY model in a transverse field, which is exactly
solvable in one dimension \cite{BarouchMD71}, and
whose entanglement properties have been the subject
of several recent investigations \cite{OsborneN02,Osterlohetal02,
Vidaletal03, Verstraeteetal04}. When $\Delta_z \neq 0$, 
and/or in higher dimensions, the model is no longer
exactly solvable. The one-dimensional case has been
indeed investigated within approximate analytical and numerical
approaches \cite{Dmitrievetal02,CapraroG02,Cauxetal03}.
A renewed interest in the model stems 
from the experimental in-field studies on the quantum spin 
chain compound Cs$_2$CoCl$_4$ \cite{Kenzelmannetal02}, 
displaying a strong planar anisotropy, $\Delta_y \approx 0.25$, 
$\Delta_z \approx 1$, and $J\approx 0.23$ meV.

\begin{figure}
\null\hspace{-.8cm}
\includegraphics[
bbllx=0pt,bblly=0pt,bburx=450pt,bbury=500pt,%
     width=55mm,angle=0]{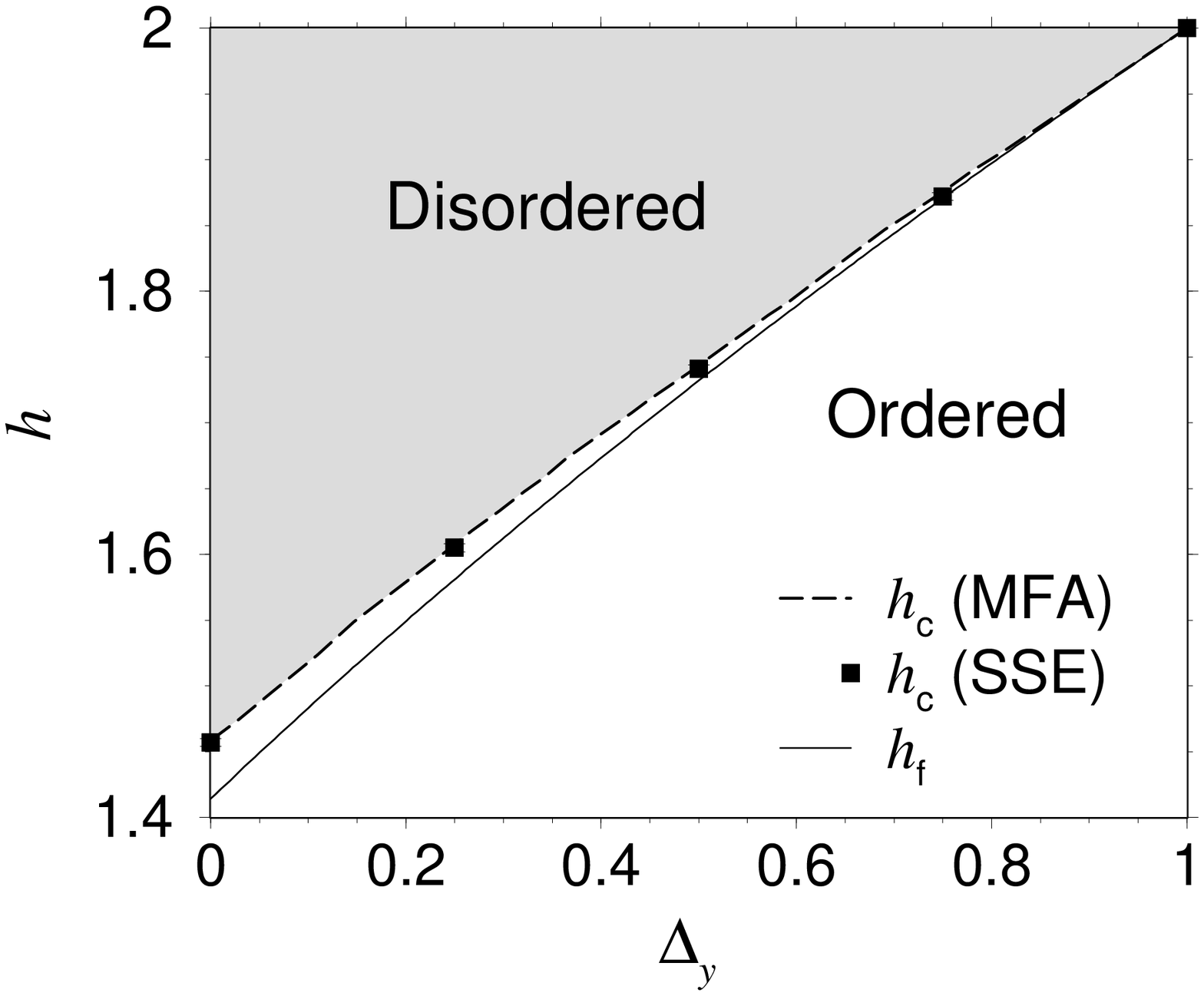} 
~~~~~~~~~~~~ \includegraphics[
bbllx=0pt,bblly=-22pt,bburx=450pt,bbury=500pt,%
     width=51mm,angle=0]{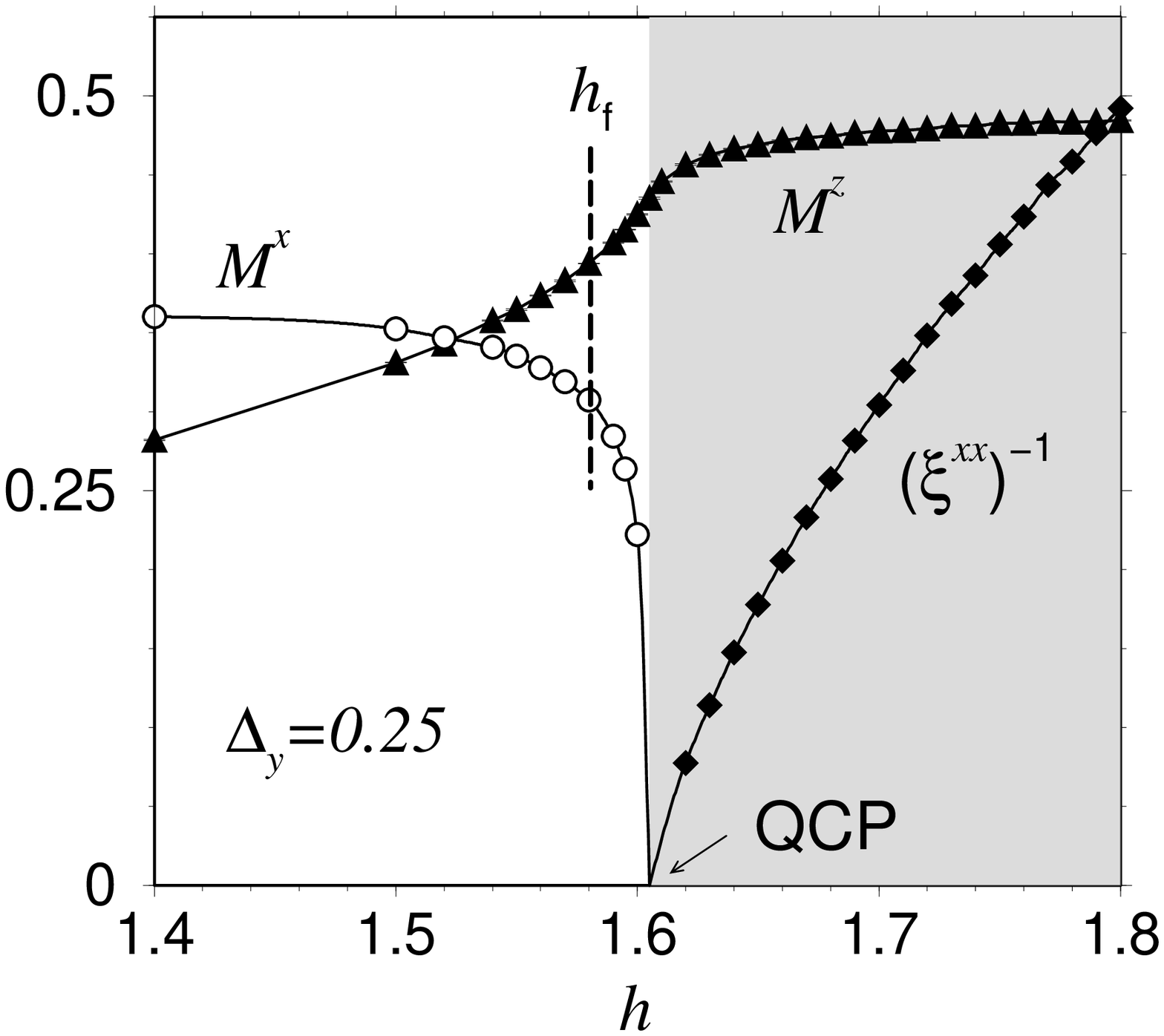}    
\caption{\label{f.XYXphd} \emph{Upper panel}. Ground state phase 
 diagram of the
 XYX model in a field. Mean-field (MFA) results are taken 
 from  Ref. \cite{Cauxetal03}, the factorizing field
 $h_{\rm f}$ from Ref. \cite{KurmannTM82}. 
 \emph{Lower panel}. Quantum critical
 behavior of $x-$ and $z-$magnetizations and 
 correlation length for
 the model with $\Delta_y=0.25$, $L=100$, $\beta=200$.
 The factorizing field is indicated by a dashed line.
 The arrow indicates the quantum critical point (QCP).
 }%
 \null\vskip -0.7cm
\end{figure} 

\section{Linear chain: critical point and factorized state}

 Motivated by the existing theoretical and experimental results, 
 we first concentrate on the case of a linear chain, and 
on the parameter range $0 \leq \Delta_y \leq 1$, 
$\Delta_z = 1$, which defines the {\it XYX model} 
in a field \cite{DelicaL90}. The qualitative behavior
of the more general XYZ model is expected to be
close to that of the XYX model in a field, since the
two models share the same symmetries. 

 The analysis of the theoretical model is performed
via Stochastic Series Expansion (SSE) Quantum
Monte Carlo (QMC) simulations \cite{SyljuasenS02},
based on a modified version of the directed-loop
algorithm to account for the low symmetry of 
the Hamiltonian \cite{Roscildeetal04}. 
Chains of various lengths, $L=40,...,120$, have been
considered, at an inverse temperature $\beta=2L$
high enough to mimic the $T=0$ behaviour for each 
lattice size.

 The left panel of Fig.~\ref{f.XYXphd} shows the ground-state phase 
 diagram of the one-dimensional XYX model in the 
 $\Delta_y-h$ plane. The quantum Monte Carlo data
 confirm that the transition belongs to the 
 universality class of the 1D transverse-field 
 Ising model (or of the 2D Ising model) 
 \cite{Dmitrievetal02}, and are in very good 
 agreement with  
 predictions from a mean-field treatment
 of the Hamiltonian \cite{Dmitrievetal02}.
 On the right panel of Fig.~\ref{f.XYXphd} we
 can clearly see the signatures of 
 the transition in the critical behavior of 
 the magnetizations along $x$ 
 (estimated through the asymptotic value of the
 spin-spin correlator as
 $M^x = |\langle \hat{S}_i^x \hat{S}_{i+L/2}^x \rangle|^{1/2}$)
 and along $z$, and the divergence of the correlation 
 length along $x$.
 
  Remarkably, none of these standard magnetic observables
  bears signatures of the second striking feature
  of the model, namely the occurrence of an exactly factorized
  state for a field $h_{\rm f}(\Delta_y)$ 
 lower than the critical field $h_{\rm c}$. The {\it factorizing field}
 reads $h_{\rm f} = \sqrt{2(1+\Delta_y)}$ in the case
 of the XYX model.  The factorized state has the 
 form $|\Psi\rangle = \bigotimes_{j=1}^{N} |\psi_j\rangle $
 with $|\psi_j\rangle = \cos\theta |\uparrow\rangle + 
 \sin\theta e^{i\phi_j} |\downarrow\rangle$  
 and $\phi_j = (1+(-1)^j)\pi/2$, 
 $\theta = \cos^{-1}\sqrt{(1+\Delta_y)/2}$. 
 This corresponds to a configuration in which the
 spins have perfectly staggered $x$ components but also
 cant out of the $xy$ plane by an angle $\theta$. 
The occurrence of such a factorized ground state
is particularly surprising if one considers that we are dealing
with the $S=1/2$ case, characterized by the most 
pronounced effects of quantum fluctuations. 
However, in the class of models here considered, 
such fluctuations are fully uncorrelated 
\cite{KurmannTM82} at $h=h_{\rm f}$, 
thus leading to a classical-like ground state.

\section{Entanglement estimators} 

 The study of entanglement
properties in the XYX model turns out to be very insightful, 
and it has the unique feature of unambiguously 
detecting both the factorized state and the quantum critical point.
 Our study has focussed on the {\it entanglement of formation} 
\cite{Bennettetal96} which is quantified through the 
{\it one-tangle} and the {\it concurrence}. 
The one-tangle \cite{Coffmanetal00,Amicoetal04} 
is an estimate of the $T=0$ entanglement between 
a single site and the remainder of the system. It is
defined as $\tau_1 = 4 \det \rho^{(1)}$, 
where $\rho^{(1)} = (I + \sum_\alpha
M^{\alpha} \sigma^{\alpha})/2$
is the one-site reduced density matrix,
$M^{\alpha} = \langle \hat{S}^{\alpha} \rangle$,
$\sigma^{\alpha}$ are the Pauli matrices, and $\alpha=x,y,z$.
In terms of the spin expectation values $M^{\alpha}$,
$\tau_1$ takes the simple form:
\begin{equation}
\tau_1 = 1 - 4 \sum_\alpha (M^{\alpha})^2 .
\end{equation}
 The concurrence \cite{Wootters98} quantifies instead
the pairwise entanglement between two spins at sites
$i$, $j$ both at zero and finite temperature.
For the model of interest, in absence of
spontaneous symmetry breaking ($M^x = 0$) the 
concurrence takes the form \cite{Amicoetal04}
\begin{equation}
C_{ij}= 2~{\rm max}\{0,C_{ij}^{(1)},C_{ij}^{(2)}\}~,
\label{e.tauC}
\end{equation}
where
\begin{eqnarray}
C_{ij}^{(1)} &=&g_{ij}^{zz}-\frac{1}{4}+|g_{ij}^{xx}-g_{ij}^{yy}|~,
\label{e.C1}\\
C_{ij}^{(2)}&=&|g_{ij}^{xx}+g_{ij}^{yy}|-
\sqrt{\left(\frac{1}{4}+g_{ij}^{zz}\right)^2-(M^z)^2}~,
\label{e.C2}
\end{eqnarray}
with $g_{ij}^{\alpha\alpha}=
\langle\hat{S}_i^\alpha\hat{S}_{j}^\alpha\rangle$. 
 \begin{figure}
\vskip -.0cm
 \begin{center}
\includegraphics[bbllx=0pt,bblly=0pt,bburx=550pt,bbury=500pt,%
     width=65mm,angle=0]{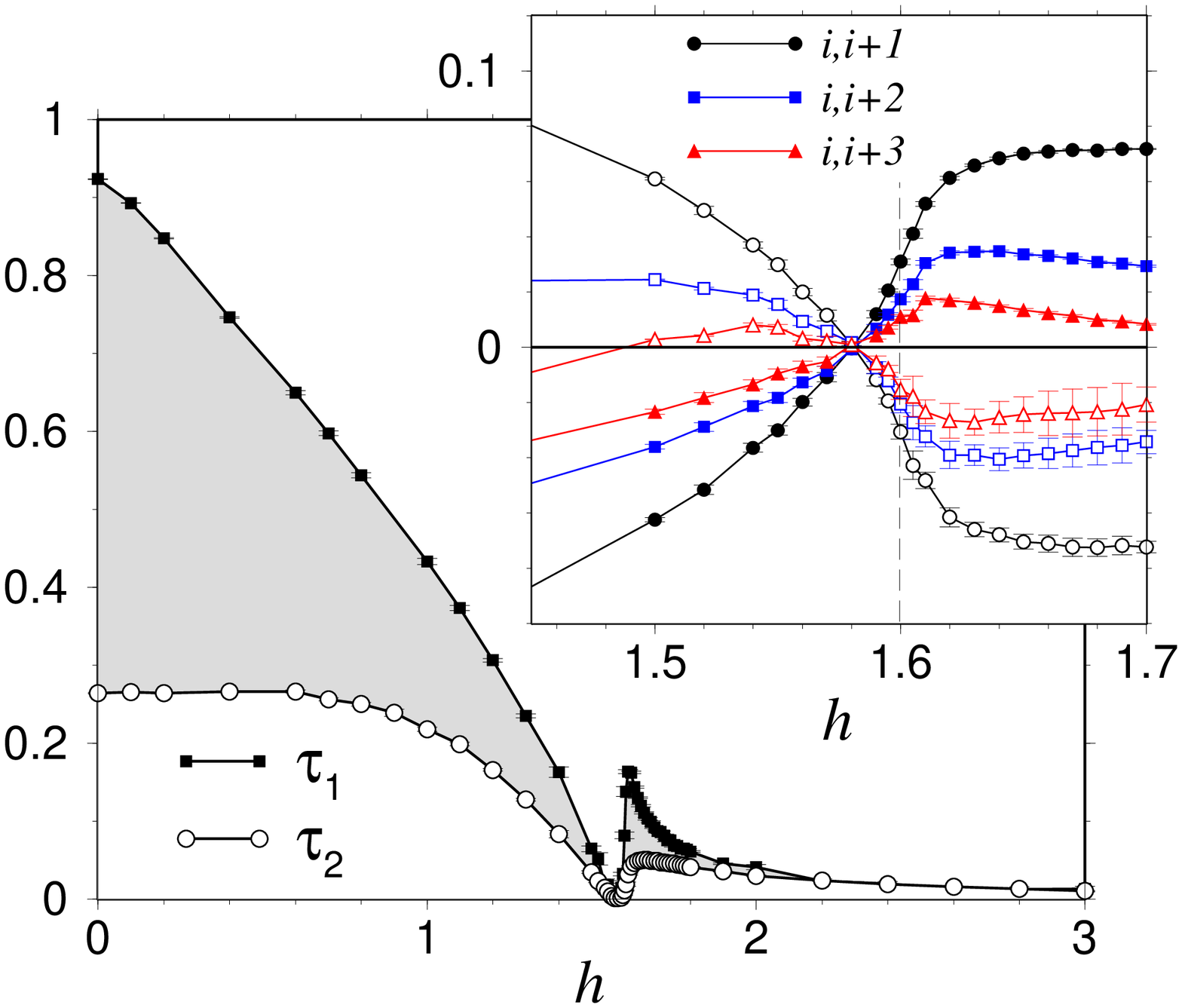}\hspace{.4cm}
\includegraphics[bbllx=0pt,bblly=0pt,bburx=550pt,bbury=500pt,%
     width=70mm,angle=0]{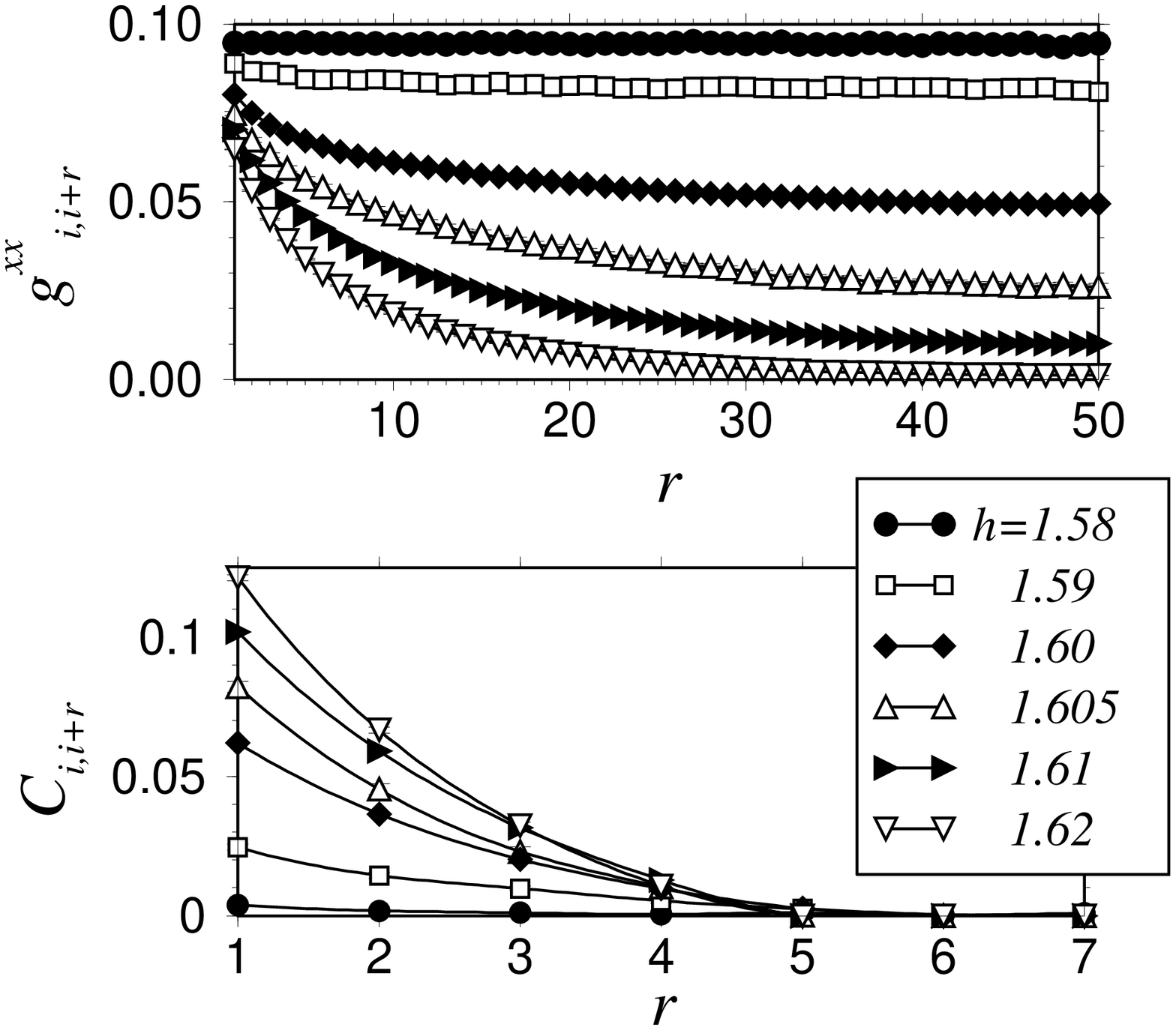}   
 \vskip -.5cm  
 \caption{\label{f.conc}\emph{Upper panel}. 
 One-tangle $\tau_1$ and sum of squared
 concurrences $\tau_2$ as a function of the applied field 
 for the $S=1/2$ XYX model with 
 $\Delta_y = 0.25$, $L = 100$ and $\beta=200$. 
 Inset: contributions to the concurrence
 between j-th neighbors; full symbols stand for 
 $C^{(1)}_{i,i+j}$, open symbols for $C^{(2)}_{i,i+j}$.
 The dashed line marks the critical field $h_{\rm c}$.
 \emph{Middle and lower panel}. Spin-spin correlator $g^{xx}_{i,i+r}$ 
 compared with the concurrence $C_{i,i+r}$ 
 for various field values
 around the critical value $h_c = 1.605(2)$. Other simulation
 parameters as in the upper panel.}
 \end{center}
\end{figure}
 \section{Results}
 The QMC results for the model Eq.~(\ref{e.XYZhz}) with
$\Delta_y = 0.25$ are shown in Fig.~\ref{f.conc},
where we plot $\tau_1$, the sum of squared concurrences
 \begin{equation}
 \tau_2 = \sum_{j\neq i} C_{ij}^2~,
 \end{equation}
 and, in the inset, $C_{i,i+n}$ for $n=1,2,3$.
 The following discussion, although directly referred to the
 results for $\Delta_y = 0.25$, is actually
 quite general and applies to all the other studied
 values of $\Delta_y$.  
 
 Unlike the standard magnetic observables plotted in 
 Fig. \ref{f.XYXphd}, the entanglement estimators display
 a marked anomaly at the factorizing field, where they 
 clearly vanish as expected for a factorized state.
 In particular, the power of the entanglement estimators
 is to rigorously unveiling the factorized nature
 of the ground state. It can indeed be easily shown that 
 the ground state is factorized \emph{if and only if} the 
 one-tangle $\tau_1$ vanishes for each spin. 
 When the field is increased above $h_{\rm f}$, the
 ground-state entanglement 
 has a very steep recovery, accompanied by the quantum
 phase transition at $h_{\rm c} > h_{\rm f}$.
 The system realizes therefore an interesting 
 {\it entanglement switch} effect controlled by the 
 magnetic field. 
 
 As for the concurrence terms Eqs.~(\ref{e.C1}),(\ref{e.C2}),
 the factorizing field divides two 
 field regions with different expressions for the
 concurrence:
 \begin{eqnarray}
 C_{ij}^{(1)}<&0&<C_{ij}^{(2)}~~~{\rm for}~~~h<h_{\rm f}~,\\
 C_{ij}^{(2)}<&0&<C_{ij}^{(1)}~~~{\rm for}~~~h>h_{\rm f}~,
 \end{eqnarray} 
 whereas $C_{ij}^{(1)} = C_{ij}^{(2)} = 0$
 at $h=h_{\rm f}$.
 In presence of spontaneous symmetry breaking
 occurring for $h < h_{\rm c}$, the expression of the 
 concurrence is generally expected to change with respect
 to Eqs.~(\ref{e.C1}),(\ref{e.C2}), as extensively discussed 
 in Ref. \cite{Syljuasen03} . For the model 
 under investigation, the expression of the concurrence 
 stays unchanged when the condition 
 \cite{Syljuasen03} $C_{ij}^{(2)} < C_{ij}^{(1)}$  
 is satisfied, i.e. for $h > h_{\rm f}$. This means
 that our estimated concurrence is accurate even in the 
 ordered phase above the factorizing field; in the 
 region $0 < h < h_{\rm f}$ it represents instead a lower
 bound to the actual $T=0$ concurrence. 
 Alternatively it can be regarded as the concurrence for 
 infinitesimally small but finite temperature. 
   
  A naive analogy between classical and quantum
correlations would lead to the expectation that the range
of pairwise entanglement, expressed through the concurrence,
is critically enhanced at a quantum phase transition, 
reflecting the divergence of the length scale for 
quantum effects. Indeed this is clearly \emph{not} the case
as shown in the right panel of Fig.~\ref{f.conc}, where 
the behavior of the spin-spin correlator $g^{xx}$
as a function of the distance
is contrasted with that of the concurrence. We observe 
that, while the correlator becomes long-ranged below
the critical field and even completely flat at the 
factorizing field, the concurrence remains short-ranged 
when passing through the transition, and it basically
vanishes after four lattice spacings. 

 \begin{figure}
 \begin{center}
\includegraphics[bbllx=0pt,bblly=0pt,bburx=550pt,bbury=380pt,%
     width=80mm,angle=0]{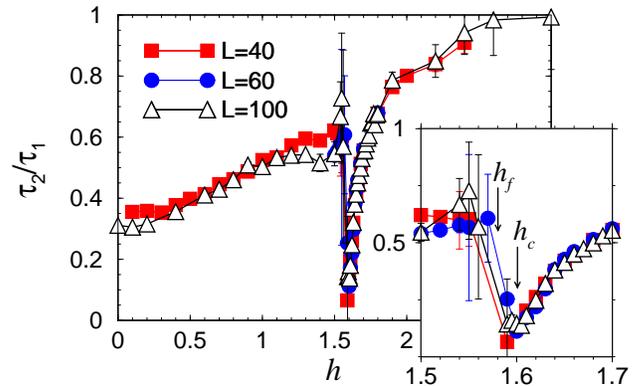}
 \caption{\label{f.ratio} Entanglement ratio $\tau_2/\tau_1$
 as a function of the field for $\Delta_y = 0.25$
 and $\beta=2L$. Inset: zoom on the critical region.}
 \end{center}
\end{figure} 

 To clarify this issue, one has to consider the special
 nature of quantum correlations. At variance 
 with classical correlations, entanglement 
 has basically to satisfy a sum rule, or, as it is 
 often phrased, is subject to a constraint of
 \emph{monogamy}. This means that the more partners
 a spin is entangled with, the less entanglement will
 be shared with each partner. Moreover, entanglement is
 present not only in the form of pairwise correlations
 as in the case of the concurrence, but also in the 
 form of $n$-spin correlations with $n>2$, which, unlike 
 classical correlations, can be 
 completely independent of pairwise correlations. This is 
 exemplified for instance by the maximally
 entangled Greenberger-Horne-Zeilinger (GHZ) state for $n$ spins 
 $|{\rm GHZ}\rangle = (|\uparrow\uparrow...\uparrow\rangle + 
 |\downarrow\downarrow...\downarrow\rangle)/\sqrt{2}$,
 on which the concurrence between any two spins is vanishing
 \cite{Coffmanetal00}. Therefore, according to monogamy,
 pairwise entanglement and $n$-wise entanglement with $n>2$
 are mutually exclusive, a condition which is absurd from 
 the point of view of classical correlations. 
 
  The mathematical expression of the monogamy constraint
  for the entanglement is provided by the Coffman-Kundu-Wootters
  (CKW) conjecture \cite{Coffmanetal00}, which indeed represents
  an approximate sum rule for entanglement
  correlations. Such conjecture states that $\tau_2 \leq \tau_1$,
  and, although it can be rigorously proved only in the 
  case of three spins, it has always been verified so far 
  in the case of an arbitrary number of spins \cite{Amicoetal04}.
  Indeed Fig. \ref{f.conc} shows that the CKW conjecture is
  verified also in the case of our model.   
  
 In particular, we interpret the {\it entanglement ratio}
 $R = \tau_2/\tau_1$ as a measure of the fraction of the total 
 entanglement stored in pairwise correlations. This
 ratio is plotted as a function of the field in Fig. \ref{f.ratio}.
 As the field increases, we observe the general trend 
 of pairwise entanglement saturating the whole entanglement
 content of the system. But a striking anomaly occurs at 
 the quantum critical field $h_c$, where $R$ displays a very 
 narrow dip. According to our interpretation, this result
 shows that the weight of pairwise entanglement decreases
 dramatically at the quantum critical point in favour of 
 multi-spin entanglement. 
  Indeed, due to the monogamy constraint, multi-spin 
 entanglement appears as the only possible quantum counterpart
 to long-range spin-spin correlations occurring at a quantum
 phase transition. The divergence of entanglement range 
 has to be interpreted in the sense that, at a quantum
 phase transition, finite $n$-spin entanglement appears
 with $n\to\infty$ at the expense of the pairwise one. 
 A unique estimator for the $n$-spin
 entanglement with $n>2$ has not been found yet, so that 
 justifying the above statement on a more quantitative
 level is problematic. Nonetheless strong indications
 of the relevance of $n$-spin entanglement with large $n$
 at a quantum critical point are given in Ref. \cite{Vidaletal03}.
 Finally, the above results evidence the 
 serious limitations of concurrence
 as a reliable estimate of entanglement at a quantum 
 critical point. 
 In turn, we propose the minimum of the entanglement 
 ratio $R$ as a novel estimator of the quantum critical
 point, fully based on entanglement quantifiers.
 
\begin{figure}
\begin{center}
\includegraphics[bbllx=0pt,bblly=0pt,bburx=550pt,bbury=450pt,%
     width=80mm,angle=0]{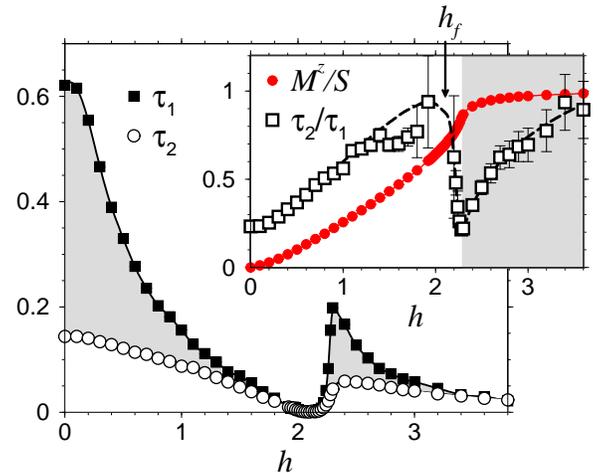}
 \caption{\label{f.ladder} One-tangle and sum of squared
 concurrences as a function of the field 
 for the XYX model on the two-leg ladder 
 with $L=40$ on each leg and $\Delta_y = 0$. 
 Inset: entanglement ratio and $z$-magnetization 
 as a function of the field. The shaded area marks the 
 quantum disordered region. All lines are guides to the eye.}
 \end{center}
\end{figure}

 \section{Two-leg ladder}
 We extend the above analysis of entanglement to the less 
 investigated case of the XYX model on a two-leg ladder, thanks
 to the fact that the SSE quantum Monte Carlo approach is 
 easily generalizable to any bipartite lattice. In this case
 the hamiltonian of Eq.(\ref{e.XYZhz}) is still expected to 
 display a field-driven quantum phase transition, whose    
 existence is independent of the lattice geometry and 
 it is instead uniquely due to the non-commutativity between
 the exchange term and the Zeeman term. There is instead
 no specific reason to expect that the model still
 displays a perfectly factorized state for geometries
 other than that of the linear chain. 
 
  Fig. \ref{f.ladder} shows the one-tangle and the 
  sum of squared concurrences for the XYX model on the 
  two-leg ladder in the case $\Delta_y = 0$. We notice
  that, for such a strong anisotropy, the two-leg
  ladder does not display a gapped Haldane phase,
  as shown by the finite value of the magnetization 
  for low fields (inset of Fig. \ref{f.ladder}).  
  Remarkably, 
  the qualitative behavior of both $\tau_1$ and 
  $\tau_2$ as 
  a function of the field is the same as in the case
  of a single chain, and in particular the vanishing
  of the one-tangle signals the rigorous existence of 
  a factorized state for a lower field than the 
  critical one. The CKW conjecture is verified for all
  the field values considered, and therefore the 
  entanglement ratio $\tau_2/\tau_1$ can be still 
  interpreted as a measure of the fraction of 
  entanglement stored in pairwise correlations. 
  This quantity, plotted in the inset of Fig. \ref{f.ladder},
  again displays a deep minimum corresponding to 
  the inflection point of the uniform magnetization and
  therefore marking the quantum critical point.
 
 \section{Conclusions}
 
  In this paper we have shown that entanglement estimators
 give a precious and novel insight in the ground state properties
 of lattice $S=1/2$ spin systems. In the case of anisotropic
 spin chains with a field-driven quantum phase transition,
 we have shown that  
 the quantum critical point can be detected through 
 a narrow dip in 
 the pairwise-to-global entanglement ratio.
 Moreover, 
 unlike the more conventional magnetic observables,
 entanglement estimators are able to single out
 the occurrence of an exactly factorized state
 in these systems. The use of quantum Monte Carlo 
 techniques naturally allows to extend this analysis 
 to different lattice geometries. In particular 
 we have discussed here the case of a two-leg ladder, 
 in which the calculation of entanglement 
 estimators remarkably shows the existence
 of a factorized state below the quantum
 critical point. The entanglement ratio displays
 again a minimum at the critical field, 
 confirming the generality of this feature
 of the entanglement behavior as a signature of 
 a quantum phase transition.
 Finally, the proximity of a
 quantum critical point to the factorized state of the
 system gives rise to an interesting field-driven 
 entanglement-switch effect. This suggests that 
 many-body effects, driven by a macroscopic parameter as
 an applied field, are a powerful tool for the control 
 of the microscopic entanglement in a multi-qubit system.
 The application of this kind of concepts in the design 
 of quantum computing devices looks therefore appealing.



\begin{acknowledgments}
 Fruitful discussions with L. Amico, T. Brun, P. Delsing, 
G. Falci, R. Fazio, A. Osterloh, and G. Vidal
are gratefully acknowledged. We acknowledge support
by DOE under grant  DE-FG03-01ER45908 (T.R. and S.H.),
by INFN, INFM, and MIUR-COFIN2002 (A.F., P.V., and V.T.).
\end{acknowledgments}



%




\end{document}